\documentstyle[11pt,aaspp4]{article}

\begin{document}


\title{Charge Transfer Between Neutral Atoms and Highly 
Ionized Species: \\ Implications for ISO Observations}

\author{
G.~J.~Ferland, K.~T.~Korista, D.~A.~Verner
}
\affil{University of Kentucky, Physics \& Astronomy, Lexington, 
KY 40506}
\author{A.~Dalgarno}
\affil{Harvard-Smithsonian Center for Astrophysics, Cambridge, MA 02138}

\begin{abstract}
We estimate rate coefficients for charge transfer between neutral
hydrogen and helium and moderate to highly ionized heavy elements. 
Although charge
transfer does not have much influence on hot collisionally ionized
plasmas, its effects on photoionized plasmas can be profound.  We
present several photoionization models which illustrate the significant
effect of charge transfer on the far infrared lines detected by ISO.

\end{abstract}


\section{Introduction}
The ionization balance of the elements in an ionized gas is determined
by photoionization and particle impact ionization and radiative,
dielectronic and three-body recombination. If the gas contains a
neutral component, charge transfer recombination may occur. In a high
temperature plasma in which ionization is produced mainly by thermal
collisions of electrons, charge transfer is rarely of significance for
highly ionized species because the abundance of neutral systems is very
small. However, in plasmas in which ionization is produced by high
frequency photons, a significant abundance of neutral hydrogen and helium
atoms may co-exist with multiply-ionized heavy elements.

\section{Charge transfer}
Because many states are energetically accessible for ions of high
nuclear charge colliding with neutral atoms, it can be argued that
electron capture by the ion is inevitable, occurring on every
collision.  The electron transfer takes place at distances where the
interaction is determined by the polarization of the atom by the
electric field generated by the approach of the ion.

If we assume that every collision that surmounts the centrifugal
barrier leads to a reaction, we obtain the Langevin formula
\begin{equation} \sigma = \pi \zeta(2\alpha/E)^{1/2}a_{o}^2
\end{equation} where $\zeta$ is the excess nuclear charge, $\alpha$ is
the dipole polarizability, $E$ is the energy of relative motion, and
$\alpha$ and $E$ are measured in atomic units. The corresponding rate
coefficient $k$ is \begin{equation} k = 3.85 \times 10^{-8} \zeta
(\alpha m/\mu)^{1/2}~\rm{cm}^3\rm{s}^{-1}, \end{equation} where $\mu/m$
is the reduced mass expressed in terms of the electron mass, $m$. For
ions of iron interacting with atomic hydrogen 
\begin{equation} 
k = 1.92 \times 10^{-9} \zeta~\rm{cm}^3\rm{s}^{-1}, 
\end{equation} 
and with atomic helium 
\begin{equation} 
k = 0.54 \times 10^{-9} \zeta~\rm{cm}^3\rm{s}^{-1}.
\end{equation} 
Figure~1
compares Equation (3) with the available data for hydrogen charge 
transfer with the first four
ionization stages of the thirty lightest elements compiled by Kingdon
\& Ferland (1996). The figure indicates the means and the sample
standard deviations of these points. Expression (3) is an upper limit
that should become more accurate as $\zeta$ increases.

\section{Applications}
We consider the abundance of Fe$^{5+}$ in a collisionally-ionized
plasma.  The temperatures at which Fe$^{5+}$ is the most abundant
ionization stage of iron lie between $1\times 10^5~\rm{K} \; \rm{and}
\; 2 \times 10^5$~K. At $10^5$~K, the rate coefficient for radiative
and dielectronic recombination of Fe$^{5+}$ is $1 \times
10^{-11}$~cm$^3$~s$^{-1}$ (Arnaud \& Raymond 1992) and the rate
coefficient for charge transfer recombination is about $1 \times
10^{-8}$~cm$^3$~s$^{-1}$. Thus, charge transfer recombination would be
comparable only if the neutral hydrogen fraction exceeded $1 \times
10^{-3}$. In equilibrium, the neutral hydrogen fraction at this
temperature is $1.7 \times 10^{-5}$ (Arnaud \& Rothenflug 1985).

The situation is very different in a cosmic plasma ionized by high
frequency photons. The photoionization cross section of atomic hydrogen
decreases with frequency $\nu$ as $\nu^{-3.5}$ (as $\nu^{-3.3}$ near
1~keV) and the photons are preferentially absorbed by heavy elements
producing a plasma in which the hydrogen is partly neutral and Auger
ionization creates multiply-ionized heavy elements.

We consider a cell of gas exposed to a blackbody radiation field at a
temperature of $\sim 10^6$~K. We adopt the photoionization cross
sections listed by Verner \& Yakovlev (1995) and Verner et al.\ (1996)
and Auger yields listed by Kaastra \& Mewe (1993). We included charge
transfer of low ionization stages of Fe, up to Fe$^{4+}$, with hydrogen
(Neufeld \& Dalgarno 1989; Kingdon \& Ferland 1996). We explored the
effects of charge transfer of more highly charged ions by carrying out
calculations with rate coefficients given by Equation (3) and with rate
coefficients set equal to zero. Figure~2 shows the distribution of
ionized stages of Fe for various values of the ionization parameter,
$U$, defined as the ratio of the ionizing photon density to the
electron density. Results for six values of $U$ are shown,
corresponding to a range from nearly neutral gas to a fully ionized
gas. Charge transfer is unimportant for the highest ionization
parameter $U$ because with it the hydrogen is almost completely
ionized, but as $U$ diminishes charge transfer causes a major
redistribution of the ionization and the abundances of the high
ionization stages are greatly reduced. The
inclusion of charge transfer with helium modifies the ionization balance
by 1\% only, due to the smaller rate coefficients and lower abundance
compared to hydrogen.
We note that a factor of 2
uncertainty in the charge transfer rate coefficient in Equations (3)
and (4)
will translate into a factor of $\sim 2^n$ uncertainty in the ionic
fractions of those ions of excess charge $\zeta = n + 5$, $n \geq 1$.

To illuminate the astrophysical impact of this charge transfer
process, we have looked at the infrared lines recently detected in
planetary nebula NGC~6302 using the Infrared Space Observatory
(Pottasch et al.\ 1996).  The observed intensities of lines [Ne II],
[Ne III], [Ne V], [Ne VI], [Mg V], [Mg VI], [Mg VIII], [Ar II], [Ar V]
were modeled by Pottasch et al.\ (1996) using the photoionization code
``Cloudy'' (Ferland 1993).  We applied a model described by Pottasch
et al.\ but included the new charge transfer recombination rates. While
the low ionization lines were not affected, some of the high ionization
lines changed considerably.  The theoretical [Mg V] intensity increased
by a factor of 1.7, whereas  the [Mg VII] and [Mg VIII] intensities
decreased by factors 2.5 and 4.0, respectively. The [Ne VI] and [Na VI]
line intensities were reduced by factors of 1.2 and 1.5, respectively,
and the [Ar V] line intensity was enhanced by a factor of 2.5. The
predicted intensity of the strongest line in the spectrum, [Ne V]
14.32~$\mu$m was virtually unaltered.  New models that include charge
transfer to highly ionized systems are needed to derive the abundances
from the observed high ionization lines.

\acknowledgments
This project was motivated by Tino Oliva's observation of anomalous
ionization of iron behind the hydrogen ionization front in
photoionization models.  We thank him for bringing this to our
attention.  One of the authors (AD) is grateful to R.\ K.\ Janev and
C-D.\ Lin for discussions.  Research in Nebular Astrophysics at the
University of Kentucky is supported by the NSF and NASA.  AD is
supported by the Division of Astronomical Science of the NSF.

\clearpage
\begin{figure}
\plotfiddle{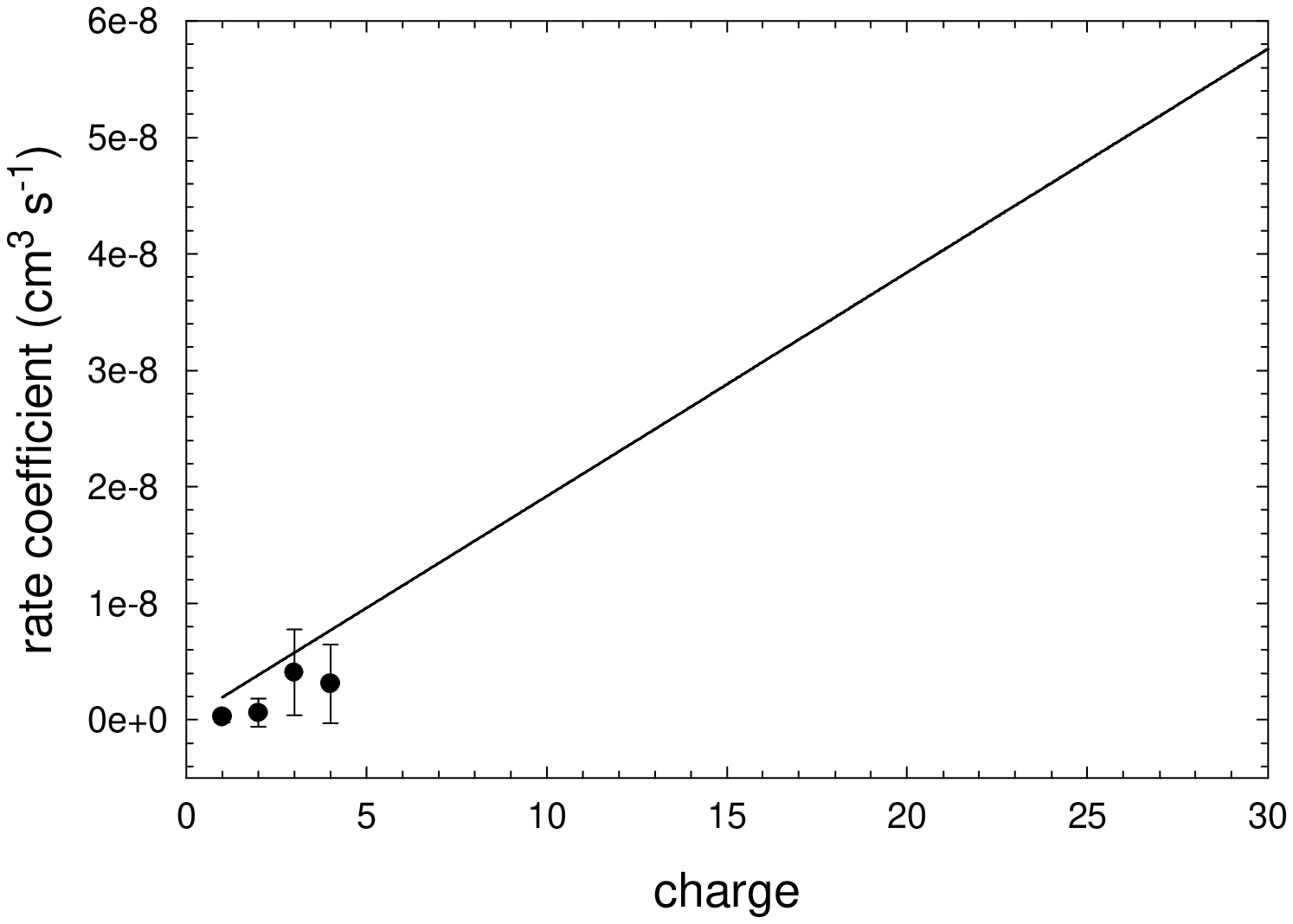}{12cm}{0}{100}{100}{-160}{250}
\caption{
Comparison between the rate 
coefficients for charge transfer with hydrogen.
The line indicates the statistical rate coefficient we adopt for
species more than four times charged.  The points indicate the mean and
standard deviation for charge transfer rate coefficients between hydrogen 
and the
first thirty elements, as given in Kingdon \& Ferland (1996).
}
\end{figure}

\clearpage
\begin{figure}
\plotfiddle{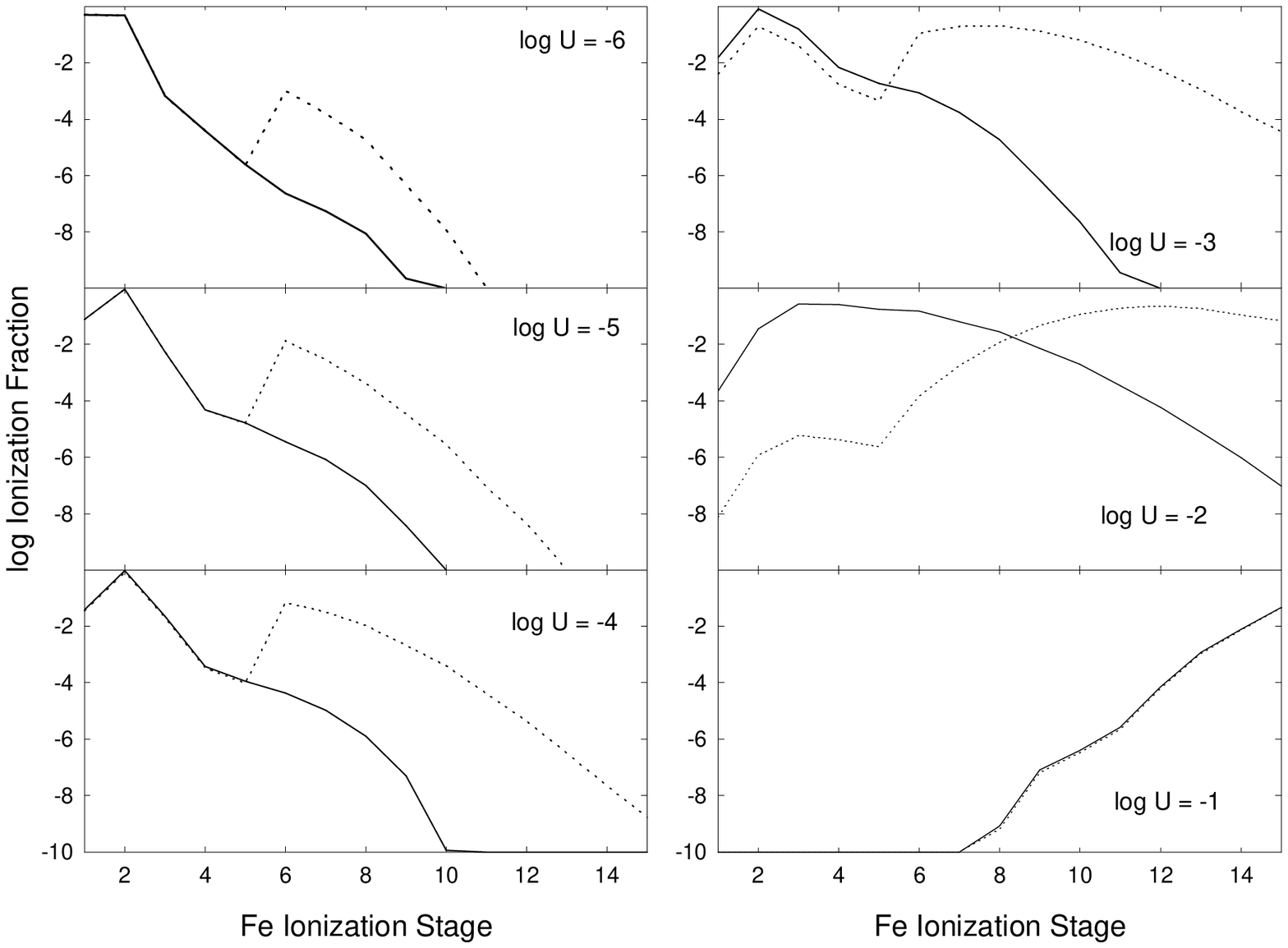}{12cm}{0}{100}{100}{-160}{250}
\caption{
Ionization distributions for iron.  Each panel shows
results for the ionization parameter indicated, photoionization by a
$10^{6.5}$~K blackbody, both with (solid) and without (dotted) the
charge transfer rates from Equation (3).
}
\end{figure}

\end{document}